\documentclass[a4paper,fleqn]{cas-dc}
\usepackage[authoryear]{natbib}
\usepackage{framed,multirow}
\usepackage{url}
\usepackage{xcolor}
\usepackage{hyperref}
\usepackage{mathrsfs}
\usepackage{subfig}
\usepackage{array}
\usepackage{booktabs}

\def\c{{\bf c}}

\def\g{{\bf g}}

\def\l{{\bf l}}

\def\q{{\bf q}}

\def\x{{\bf x}}

\def\0{{\bf 0}}
\def\1{{\bf 1}}

\def\etal{{\em et al.}}
\def\eg{{\em e.g.}}
\def\ie{{\em i.e.}}
\def\etc{{\em etc}}

\def\etal{{\em et al.\/}\,}

\usepackage{threeparttable}
\usepackage{tabularx}
\usepackage{rotating}
\usepackage{color} 
\usepackage{soul}
\usepackage{flushend}

\def\tsc#1{\csdef{#1}{\textsc{\lowercase{#1}}\xspace}}
\tsc{WGM}
\tsc{QE}
\tsc{EP}
\tsc{PMS}
\tsc{BEC}
\tsc{DE}

\begin{document}
\begin{sloppypar}

\let\WriteBookmarks\relax
\def\floatpagepagefraction{1}
\def\textpagefraction{.001}
\def\etal{{\em et al.}}
\def\eg{{\em e.g.}}
\def\ie{{\em i.e.}}
\def\etc{{\em etc}}

\shorttitle{ACTION: Augmentation and Computation Toolbox for Brain Network Analysis with Functional MRI}

\shortauthors{Fang et~al.}

\title [mode = title]{ACTION: Augmentation and Computation Toolbox for Brain Network Analysis with Functional MRI}

\author[1]{Yuqi Fang}
\author[2]{Junhao Zhang}
\author[2]{Linmin Wang}
\author[1]{Qianqian Wang}
\author[1]{Mingxia Liu}[orcid=0000-0002-0166-0807]
\cormark[1]
\ead{mingxia_liu@med.unc.edu}
\cortext[cor1]{Corresponding author}

\affiliation[1]{organization={Department of Radiology and Biomedical Research Imaging Center}, addressline={University of North Carolina at Chapel Hill}, city={Chapel Hill}, state={NC 27599}, country={United States}}
\affiliation[2]{organization={School of Mathematics Science}, addressline={Liaocheng University}, city={Liaocheng}, state={Shandong 252000}, country={China}}


\begin{abstract}
Functional magnetic resonance imaging (fMRI) has been increasingly employed to investigate functional brain activity. 
Many fMRI-related software/toolboxes have been developed, providing specialized algorithms for fMRI analysis. 
However, existing toolboxes seldom consider fMRI data augmentation, which is quite useful, especially in studies with limited or imbalanced data. 
Moreover, current studies usually focus on analyzing fMRI using conventional machine learning models that rely on human-engineered fMRI features, without investigating deep learning models that can automatically learn data-driven fMRI representations. 
In this work, we develop an open-source toolbox, called \textbf{A}ugmentation and \textbf{C}omputation \textbf{T}oolbox for bra\textbf{I}n netw\textbf{O}rk a\textbf{N}alysis (\textbf{ACTION}), offering comprehensive functions to streamline fMRI analysis. 
The ACTION is a Python-based and cross-platform toolbox with graphical user-friendly interfaces. 
It enables automatic fMRI augmentation, covering blood-oxygen-level-dependent (BOLD) signal augmentation and brain network augmentation. 
Many popular methods for brain network construction and network feature extraction are included. 
In particular, it supports constructing deep learning models, which leverage large-scale auxiliary unlabeled data (3,800+ resting-state fMRI scans) for model pretraining to enhance model performance for downstream tasks. 
To facilitate multi-site fMRI studies, it is also equipped with several popular federated learning strategies. 
Furthermore, it enables users to design and test custom algorithms through scripting, greatly improving its utility and extensibility. 
We demonstrate the effectiveness and user-friendliness of ACTION on real fMRI data and present the experimental results. 
The software, along with its source code and manual, can be accessed \href{https://github.com/mxliu/ACTION-Software-for-Functional-MRI-Analysis/tree/main/Software}{online}.

\end{abstract}



\begin{keywords}
Toolbox \sep Functional MRI Augmentation \sep Brain Network Analysis \sep Deep Learning Model \sep Federated Learning 
\end{keywords}
\maketitle

\section{Introduction}
Functional magnetic resonance imaging (fMRI) provides a noninvasive imaging technique for measuring spontaneous brain activity by detecting changes in blood-oxygen-level-dependent (BOLD) signals~\citep{fox2007spontaneous}. 
It has been increasingly employed to investigate functional activities of the brain, demonstrating great clinical and practical value in many applications, including neurological disease diagnosis~\citep{zhang2023gcl}, brain development assessment~\citep{edde2021functional}, and biomarker identification~\citep{hu2021multiband}.

Currently, many fMRI-related software and toolboxes have been developed~\citep{kruschwitz2015graphvar, treder2020mvpa, lanka2020malini, waller2018graphvar, treder2020mvpa, lanka2020malini, xu2018brant, meunier2020neuropycon, zhou2020toolbox}, offering specialized algorithms for users to facilitate fMRI analysis in an efficient and standardized manner. 
For example, some toolboxes~\citep{kruschwitz2015graphvar, zhou2020toolbox, wang2015gretna} focus on constructing functional connectivity networks based on fMRI and generating network topological features, which allows for identifying disease-associated brain functional alterations. 
Some other studies~\citep{waller2018graphvar, treder2020mvpa, lanka2020malini} assist in constructing machine learning models for brain disorder analysis, which can greatly enhance the efficiency of medical decision-making.

\if false
However, existing studies 
usually utilize original fMRI data for computation analysis, neglecting to perform data augmentation to expand the size and diversity of the given fMRI data. 
\fi 
However, existing studies usually utilize original fMRI data for computation analysis, ignoring the basic function of enhancing the size and diversity of given fMRI data (\ie, data augmentation). 
Functional MRI augmentation is quite useful, especially in studies with limited data samples, which can help improve the robustness and generalization of the constructed learning models.
Additionally, current works usually investigate fMRI using conventional machine learning models that rely on human-engineered fMRI features, without exploring deep learning models that can automatically learn data-driven fMRI feature representations.
Compared with machine learning methods, deep learning models typically integrate feature learning and model construction into one united model, resulting in data-driven features, which may lead to improved prediction results. 

\begin{figure*}[!t]
\setlength{\abovecaptionskip}{-0pt}
\setlength{\belowcaptionskip}{-0pt} 
\setlength\abovedisplayskip{-0pt}
\setlength\belowdisplayskip{-0pt}
\center
\includegraphics[width= 1\linewidth]{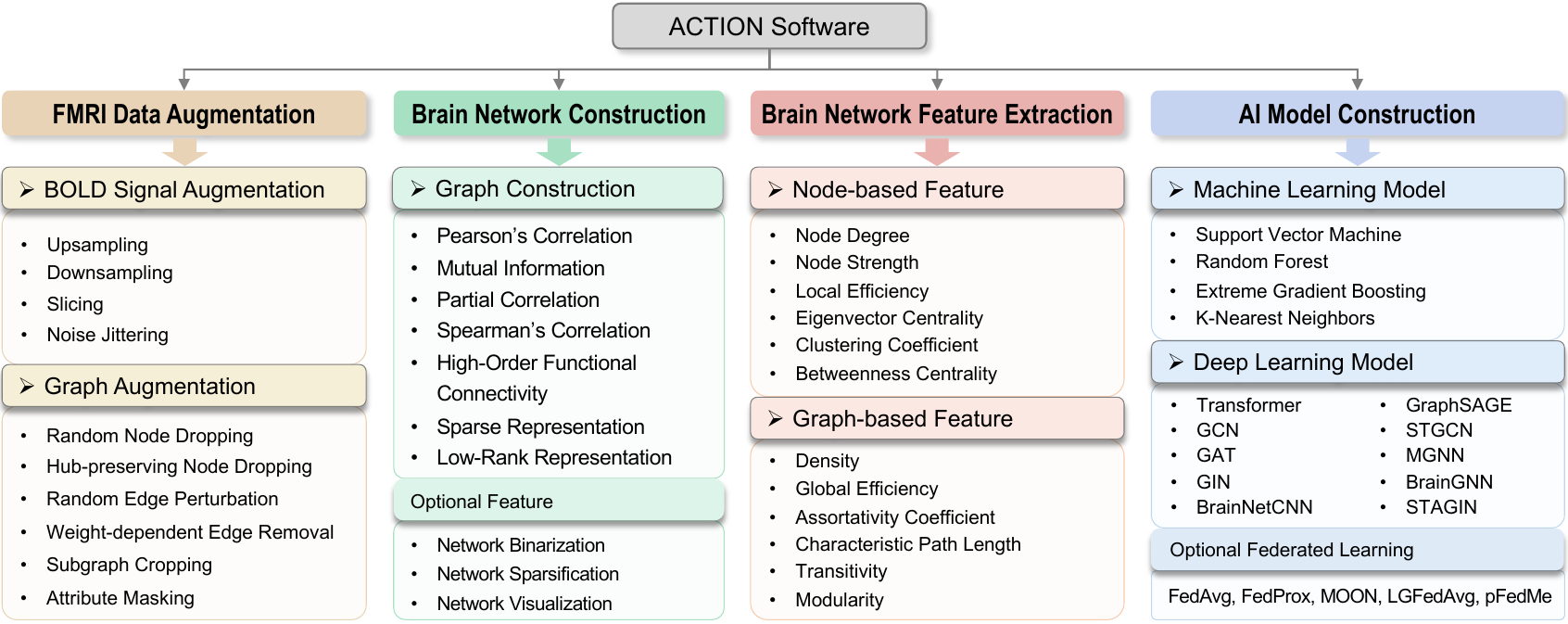}
\caption{Major functions included in the proposed ACTION software, including fMRI data augmentation, brain network construction, brain network feature extraction, and artificial intelligence (AI) model construction.}
\label{fig_pipeline}
\end{figure*}

To this end, we develop an open-source toolbox, called \textbf{A}ugmentation and \textbf{C}omputation \textbf{T}oolbox for bra\textbf{I}n netw\textbf{O}rk a\textbf{N}alysis (\textbf{ACTION}), which offers comprehensive functions to streamline fMRI analysis. 
The ACTION is a Python-based and cross-platform (Windows, Linux, and Mac OS) toolbox with graphical user-friendly interfaces, and its major functions can be found in Fig.~\ref{fig_pipeline}.
The ACTION features the following advantages compared to most existing works.
\emph{First}, it enables automatic fMRI data augmentation, including both BOLD signal augmentation and brain network/graph augmentation. 
\emph{Second}, ACTION integrates many methods for brain functional connectivity network construction and supports extracting multiple brain network features, including node-based and graph-based features.
\emph{Third}, besides machine learning models, it also supports the construction of deep learning models, where ten popular methods for fMRI analysis are embedded.
It is noteworthy that, for each method, our toolbox provides a pretrained deep learning model based on large-scale unlabeled fMRI data (3,800+ scans).
\emph{In addition}, it also integrates several popular federated learning strategies to facilitate multi-site fMRI studies. 
\if false
\emph{In addition}, ACTION integrates many methods for brain functional connectivity network construction and supports extracting multiple brain network features, including node-based and graph-based features.
These functions, alongside data augmentation and model construction, effectively streamline fMRI analysis.
\fi 
\emph{Furthermore}, it enables users to design and test their custom algorithms through scripting, which greatly improves its utility and extensibility. 
To demonstrate the effectiveness and user-friendliness of ACTION, we employ real fMRI data for model evaluation. 
Detailed comparison between our ACTION and existing toolboxes for computer-aided fMRI analysis is shown in Table~\ref{tab_toolbox_comparison}. 


The remainder of this paper is organized as follows. 
Section~\ref{sec_method} details the proposed ACTION, including all function modules and the corresponding algorithms. 
Specifically, Section~\ref{sec_data_aug} introduces two types of fMRI data augmentation.
Section~\ref{sec_network_construct} and Section~\ref{sec_network_feat_extract} present functions about brain network construction and brain network feature extraction based on fMRI data, respectively.
Section~\ref{sec_ai_model_construct} introduces artificial intelligence (AI) model construction, covering both conventional machine learning models and deep learning models.
In Section~\ref{sec_experiment_validation}, we validate the effectiveness of included models using real resting-state fMRI data and present the corresponding experimental results.
The paper is concluded in Section~\ref{sec_conclusion}.

\begin{table*}
\setlength{\abovecaptionskip}{-0pt}
\setlength{\belowcaptionskip}{-0pt} 
\setlength\abovedisplayskip{-0pt}
\setlength\belowdisplayskip{-0pt}
    \small
    \renewcommand\arraystretch{1}
    \caption{Comparison of major functions between the proposed ACTION and existing toolboxes for computer-aided functional MRI analysis.
    }
    \label{tab_toolbox_comparison}
    \centering
    \renewcommand\arraystretch{1.4}
    \rowcolors{2}{gray!15}{white}
    \resizebox{\textwidth}{!}{
    \begin{tabular}{l|cccccccc}
    \toprule
    Toolbox & \makecell[c]{Programming\\Language} & \makecell[c]{Graphical\\User Interface}  & \makecell[c]{FMRI Data\\Augmentation} & \makecell[c]{Functional Brain \\Network Construction} & \makecell[c]{Brain Network\\Feature Extraction} & \makecell[c]{Machine Learning\\Model Construction} & \makecell[c]{Deep Learning\\Model Construction}  & \makecell[c]{Federated Learning\\for Multi-site fMRI Analysis}\\
    \midrule
    
    PyMVPA~\citep{hanke2009pymvpa} & Python &$\times$&$\times$&$\times$&$\times$& \checkmark\ (Classification, Regression) &$\times$ &$\times$\\ 
    
    BCT~\citep{rubinov2010complex} & Matlab &$\times$&$\times$&$\times$& \checkmark &$\times$&$\times$ &$\times$\\
    
    REST~\citep{song2011rest} & Matlab & \checkmark &$\times$& \checkmark &$\times$&$\times$&$\times$ &$\times$\\
    
    CONN~\citep{whitfield2012conn} & Matlab & \checkmark &$\times$& \checkmark & \checkmark &$\times$&$\times$ &$\times$\\
    
    PRoNTo~\citep{schrouff2013pronto} & Matlab & \checkmark &$\times$&$\times$&$\times$& \checkmark\ (Classification, Regression) &$\times$ &$\times$\\
    
    MANIA~\citep{grotegerd2014mania} & Matlab & \checkmark &$\times$&$\times$&$\times$& \checkmark\ (Classification) &$\times$ &$\times$\\
    
    DynamicBC~\citep{liao2014dynamicbc} & Matlab & \checkmark &$\times$& \checkmark & \checkmark &$\times$&$\times$ &$\times$\\
    
    BASCO~\citep{gottlich2015basco} & Matlab & \checkmark &$\times$& \checkmark & \checkmark &$\times$&$\times$ &$\times$\\
    
    GraphVar~\citep{kruschwitz2015graphvar} & Matlab & \checkmark &$\times$& \checkmark & \checkmark &$\times$&$\times$ &$\times$\\
    
    GRETNA~\citep{wang2015gretna} & Matlab & \checkmark &$\times$& \checkmark & \checkmark &$\times$&$\times$ &$\times$\\
    
    GraphVar 2.0~\citep{waller2018graphvar} & Matlab & \checkmark &$\times$& \checkmark & \checkmark & \checkmark\ (Classification, Regression) &$\times$ &$\times$\\
    
    BRANT~\citep{xu2018brant} & Matlab & \checkmark &$\times$& \checkmark & \checkmark &$\times$&$\times$ &$\times$\\
    
    MVPA-Light~\citep{treder2020mvpa} & Matlab &$\times$&$\times$&$\times$&$\times$& \checkmark\ (Classification, Regression) &$\times$ &$\times$\\
    
    BrainNetClass~\citep{zhou2020toolbox} & Matlab & \checkmark &$\times$& \checkmark & \checkmark & \checkmark\ (Classification) &$\times$ &$\times$\\
    
    MALINI~\citep{lanka2020malini} & Matlab & \checkmark &$\times$& \checkmark &$\times$& \checkmark\ (Classification, Regression) &$\times$ &$\times$ \\
    
    NeuroPycon~\citep{meunier2020neuropycon} & Python &$\times$&$\times$& \checkmark & \checkmark &$\times$&$\times$ &$\times$ \\ 

    \textbf{ACTION (Ours)} & Python & \checkmark & \checkmark & \checkmark & \checkmark & \checkmark\ (Classification, Regression) & \checkmark & \checkmark \\
    \bottomrule
    \end{tabular}}
\end{table*}

\section{Functions of ACTION}\label{sec_method}
The ACTION software includes four major functions, \ie, fMRI data augmentation, brain network construction, brain network feature extraction, and AI model construction. 
The software and its open-source codes can be accessed via \url{https://github.com/mxliu/ACTION-Software-for-Functional-MRI-Analysis/tree/main/Software}, with four function modules detailed as follows. 

\subsection{Functional MRI Data Augmentation}\label{sec_data_aug}
Functional MRI augmentation refers to the technique that enhances the quantity and diversity of fMRI.
It usually helps improve the robustness of constructed models in fMRI analysis.
Typically, there are two mainstream methods for fMRI data augmentation, \ie, \emph{BOLD signal augmentation} and \emph{graph augmentation}. 
Here, a graph corresponds to a specific brain connectivity network derived from fMRI.

\subsubsection{BOLD Signal Augmentation}

\begin{figure}[!t]
\center
\includegraphics[width= 1\linewidth]{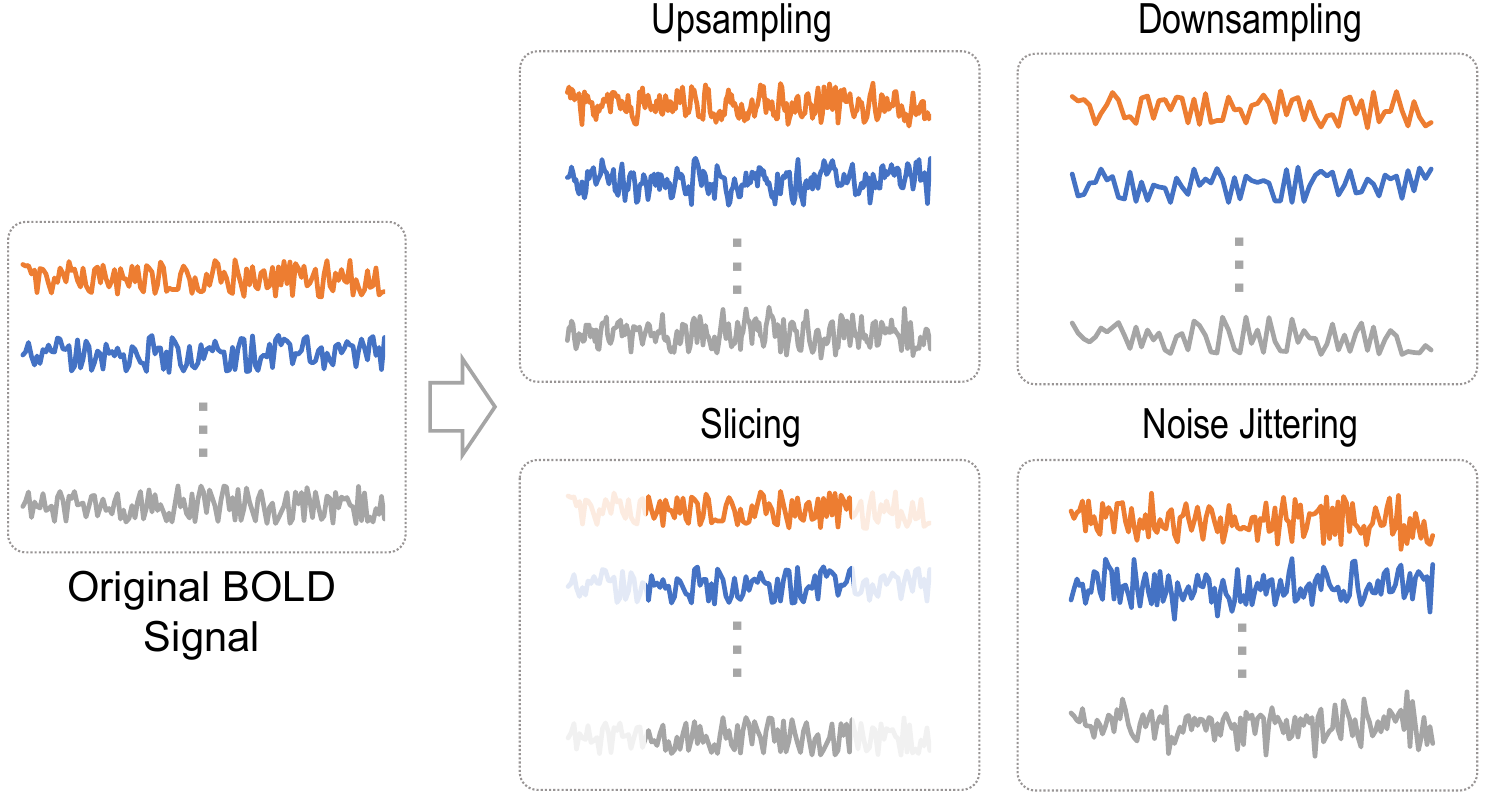}
\caption{Illustration of four methods for fMRI blood-oxygen-level-dependent (BOLD) signal augmentation.}
\label{fig_bold_aug}
\end{figure}

Many fMRI studies~\citep{li2021braingnn, dvornek2018learning, wang2023unsupervised} directly perform data augmentation based on the raw BOLD signals.
These methods focus on introducing variations to fMRI time series, which simulate various temporal dynamics of brain activity.
As shown in Fig.~\ref{fig_bold_aug}, four popular methods for fMRI BOLD signal augmentation are included, \ie, upsampling, downsampling, slicing, and noise jittering. 
In addition to these methods, our toolbox also supports users to design their own BOLD signal augmentation algorithms.
The details for custom algorithm deployment can be found in  \emph{Supplementary Materials}. 
\if false
The illustration of these methods is given in Fig.~\ref{fig_bold_aug}, and each method is described as follows. 
\fi 

1) \textbf{Upsampling} is an augmentation strategy to stretch a time series~\citep{le2016dataaug}, which increases fMRI temporal resolution and captures more rapid neural activity changes.
Specifically, given an fMRI time series with $T$ timepoints, we perform upsampling using fast Fourier transform~\citep{brigham1988fast} with ratio $u\in (0,1)$.
This results in a new fMRI time series with $\lfloor T/u\rfloor$ timepoints, where $\lfloor.\rfloor$ represents a floor function.
The newly derived data can be used for further analysis, \eg, constructing brain functional networks or building learning based models.

2) \textbf{Downsampling} aims to contract a time series by decreasing its resolution~\citep{le2016dataaug}, helping capture more coarse-grained patterns and more general temporal trends.
To perform downsampling, we leverage a fast Fourier transform using a ratio $b\in (0,1)$, resulting in an fMRI time series with $\lfloor T\times b\rfloor$ timepoints.

3) \textbf{Slicing} focuses on dividing the fMRI time series data into smaller segments~\citep{le2016dataaug}.
Each segment represents a subset of the original time series, which captures localized temporal patterns of fMRI data.
Given an fMRI with $T$ timepoints and a slicing ratio $s\in (0,1)$, we can derive a segmented fMRI with length $\lfloor T\times s\rfloor$, and the starting point of the new time series can change dynamically.

4) \textbf{Noise Jittering}~\citep{wen2021timedataaug} is to add random noise to the fMRI data, simulating real-world noise such as motion artifacts.
Here, we employ Gaussian noise to introduce randomness to fMRI time series.
Gaussian noise is a type of noise with the probability density function following the normal distribution. 
We denote Gaussian noise as $g\in\mathbb{R}^{T}$. 
Each element of $g$ is a sample from the normal distribution $N(\mu, \sigma)$, where $\mu$ and $\sigma$ denote the expectation/mean and standard deviation of the distribution.
Given an fMRI time series $X\in\mathbb{R}^{T}$ and noise $g$, we can easily generate the new time series: $X$+$g$.

\if false
In addition to these methods, our toolbox also supports users to design their own BOLD signal augmentation algorithms.
The details for custom algorithm deployment can be found in the \emph{Supplementary Materials}. 
\fi 

\begin{figure}[!t]
\setlength{\abovecaptionskip}{-0pt}
\setlength{\belowcaptionskip}{-0pt} 
\setlength\abovedisplayskip{-0pt}
\setlength\belowdisplayskip{-0pt}
\center
\includegraphics[width= 1\linewidth]{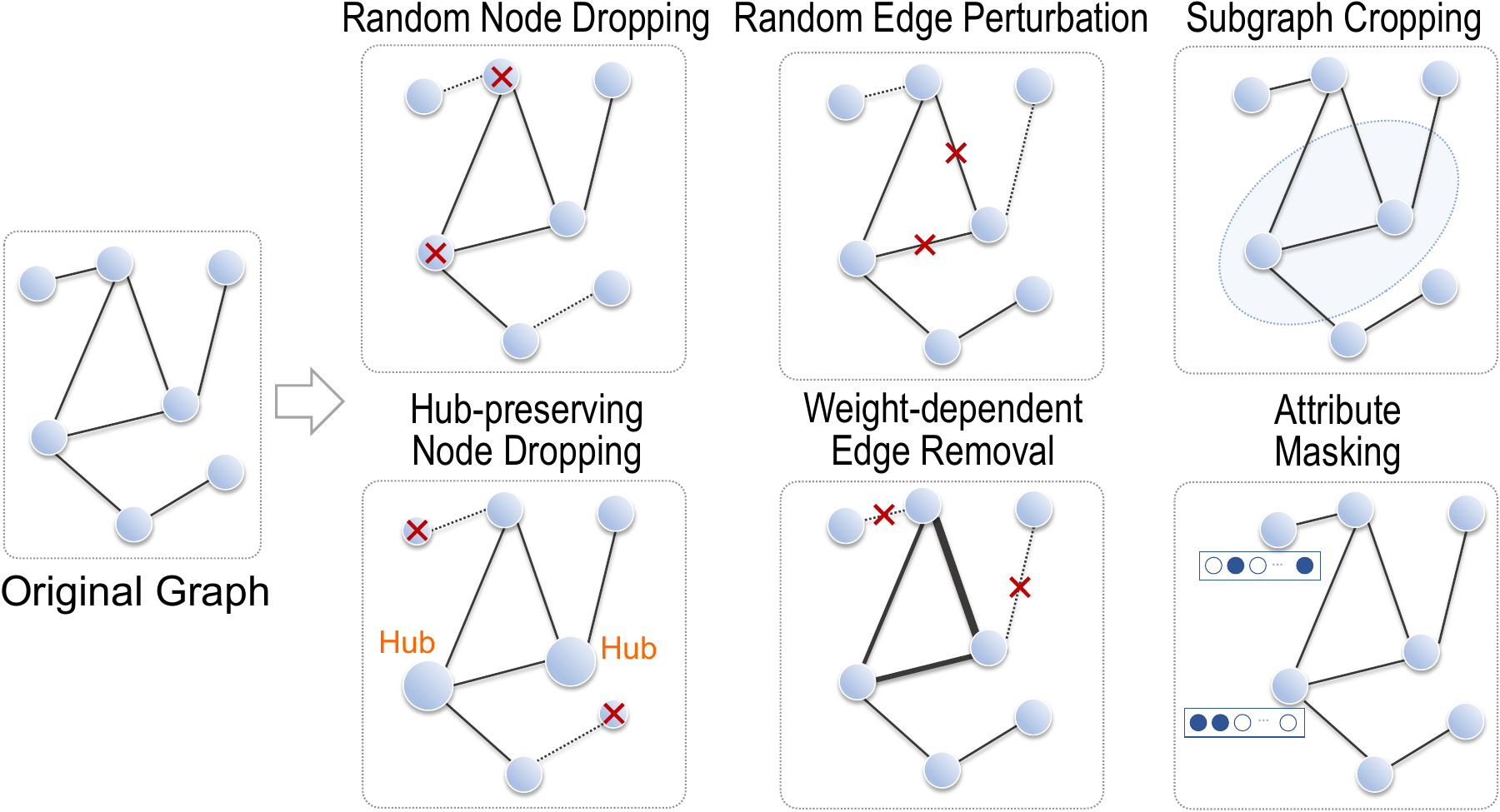}
\caption{Illustration of six graph augmentation methods based on fMRI-derived brain networks/graphs.}
\label{fig_graph_aug}
\end{figure}

\subsubsection{Graph Augmentation}

Rather than directly manipulate original fMRI time series for data augmentation, some other fMRI studies~\citep{pei2022dataaug} focus on first representing the brain as a functional connectivity network and then augmenting the brain network to increase data diversity.
Each brain network contains a set of interconnected brain regions that exhibit correlated or synchronized functional activity patterns.
These networks are often represented as \emph{graphs}, where graph nodes denote brain regions-of-interest (ROIs) and graph edges mean functional connections between paired  ROIs. 
Graph augmentation aims to introduce variability to the brain network/graphs constructed by fMRI, such as dropping nodes, perturbing edges, or modifying graph structure.
With graph augmentation, we can capture a wider range of potential functional connectivity patterns of the brain, helping facilitate fMRI data analysis and advance our understanding of brain functionality.

Denote a brain network/graph as $G=\{V, E, A\}$, where each node $v_{i}\in{V}$ represents a brain ROI with its feature vector $h_{i}\in\mathbb{R}^{D}$, and $E$ denotes the edge set. 
$A\in\mathbb{R}^{N\times N}$ is a matrix with each element $a_{ij}$ denoting functional connectivity between the $i$-th and the $j$-th ROIs, where $N$ denotes the number of nodes/ROIs.
In ACTION, we introduce six methods for graph augmentation (see Fig.~\ref{fig_graph_aug}), including four popular methods (\ie, random node dropping, random edge perturbation, subgraph cropping, and attribute masking) and two recently designed methods (\ie, hub-preserving node dropping and weight-dependent edge removal). 
  
1) \textbf{Random Node Dropping}~\citep{you2020graphcl} aims to randomly remove a certain proportion of graph nodes along with their associated edges from $G$.
The probability of dropping each node follows a uniform distribution, and the node dropping rate $o$ ranges from 0 to 1.
For instance, if $o$ is set as 0.05, it means that 5\% of graph nodes, along with the connected edges, are randomly discarded.

2) \textbf{Hub-Preserving Node Dropping} is a recently designed algorithm for graph augmentation, which prioritizes preserving brain hub regions during this procedure.
Brain hubs refer to brain regions that exhibit a high degree of functional connectivity with other regions and they usually play a significant role in facilitating effective interaction within the brain.
Here, we employ degree centrality~\citep{rubinov2010complex} to measure the importance of each node and determine the dropping probability based on its degree centrality $d_i$.
Specifically, the probability of the node $v_i$ being dropped is represented as $q_i=1/d_i$. 
In this way, the nodes with a higher degree centrality exhibit a lower dropping probability.
Then, we obtain probability distribution $p_i$ for node $v_i$ based on normalization: $p_i=q_i/\sum\nolimits_{i=1}q_i$.
Following this probability distribution, we drop certain nodes according to a specified drop ratio $c\in (0,1)$ for graph augmentation.

3) \textbf{Random Edge Perturbation}~\citep{you2020graphcl} aims to perturb the graph edges in $G$ by randomly adding or dropping a certain proportion of edges, while maintaining the total number of edges consistent with the original graph.
The probability of adding or dropping each edge follows a uniform distribution, with an edge adding/dropping ratio $e\in (0,1)$.
If $e$ is set as 0.1, 10\% edges are randomly removed, and 10\% edges are randomly added.

4) \textbf{Weight-Dependent Edge Removal}  is a graph augmentation method based on edge perturbation.
Rather than randomly remove edges, it performs edge dropping by considering edge importance/weight (\ie, $a_{ij}$).
Specifically, it uses Pearson's correlation coefficient~\citep{freedman2007statistics} to measure the edge weight $a_{ij}$, and the probability for an edge $e_{ij}$ being removed is defined as $p_{ij}=q_{ij}/\sum\nolimits_{i=1}\sum\nolimits_{j=1}q_{ij}$, where $q_{ij}=1/a_{ij}$.
That is, the edges with stronger functional connectivity have a lower probability of being dropped.
According to this probability, we remove a certain proportion of edges to generate augmented graphs based on a given dropping ratio.

5) \textbf{Subgraph Cropping}~\citep{you2020graphcl} randomly selects a subset of brain regions and their associated functional connections to create a smaller subnetwork based on a random walk.
The underlying assumption is that the semantic information of $G$ can be well preserved within its partial graph structure.

6) \textbf{Attribute Masking}~\citep{you2020graphcl} involves randomly masking attributes or features associated with certain brain regions/nodes.
It assumes that missing partial node features do not impact the whole graph much.

In addition to the above-mentioned methods, users can design and use their custom graph augmentation algorithms. 
More details on implementing self-defined graph augmentation algorithms can be found in \emph{Supplementary Materials}.

\subsection{Brain Network Construction}\label{sec_network_construct}
Our toolbox includes seven popular methods for brain network construction, \ie, \emph{Pearson's correlation}, \emph{mutual information}, \emph{partial correlation}, \emph{Spearman's correlation}, \emph{high-order functional connectivity}, \emph{sparse representation}, and \emph{low-rank representation}.
It also embeds two network sparsification strategies, helping focus on the strong connections representing neural processes.
Moreover, it supports brain network visualization, allowing users to identify functional connectivity patterns that may not be apparent shown in original data.
The algorithms for network construction and the details of network sparsification and visualization are detailed as follows.

\subsubsection{Network Construction Methods}
1) \textbf{Pearson's Correlation} (PC)~\citep{cohen2009pearson} measures linear relationships between fMRI time series of different brain regions.
Denote $X$ and $Y$ as fMRI time series of two brain regions, and the PC coefficient between $X$ and $Y$ can be derived using:
\begin{equation}
PC_{XY} = \frac{cov(X, Y)}{\sigma_{X} \sigma_{Y}},
\end{equation}
where $cov(X, Y)$ denotes the covariance between $X$ and $Y$, and $\sigma_{X}$ and $\sigma_{Y}$ represent the standard deviation of $X$ and $Y$, respectively.
Typically, $PC_{XY}$ ranges from $-1$ and $1$.
A value close to $1$ indicates stronger positive synchronization between $X$ and $Y$, suggesting that the corresponding brain regions are functionally connected and likely involved in similar neural processes.
And a value close to $-1$ indicates a strong negative relationship.
A value close to $0$ suggests no linear relationship between time series of paired regions. 

2) \textbf{Mutual Information} (MI)~\citep{kraskov2004estimating} quantifies the amount of information we can obtain from one fMRI time series by observing the other fMRI time series.
Mathematically, the MI between two fMRI time series $X$ and $Y$ can be represented as:
\begin{equation}
MI_{XY} = \sum\nolimits_{y \in Y} \sum\nolimits_{x \in X} p(x,y) \log \left( \frac{p(x,y)}{p(x)p(y)} \right),
\end{equation}
where $p(x,y)$ denotes the joint probability distribution function of $X$ and $Y$.
And $p(x)$ and $p(y)$ are the marginal probability distribution functions of $X$ and $Y$, respectively. 
A higher value of $MI_{XY}$ indicates a stronger dependency between two fMRI time series, indicating greater synchronization in neural activity between two brain regions.
If $MI_{XY}$ is close to $0$, it suggests that two fMRI time series are independent.
Compared with Pearson's correlation, which only measures the linear relationship between two time series, mutual information can capture both linear and nonlinear relationships between them.

3) \textbf{Partial Correlation} (PrC)~\citep{de2004discovery} examines the relationship between fMRI time series of two regions while controlling for the influence of other brain regions.
The partial correlation coefficient between two fMRI time series $X$ and $Y$ can be denoted as follows:
\begin{equation}
PrC_{XY} = \frac{\rho_{XY} - \rho_{XZ} \rho_{YZ}}{\sqrt{(1 - \rho_{XZ}^2)(1 - \rho_{YZ}^2)}},
\end{equation}
where $Z$ represents the time series of all the other remaining regions.
$\rho_{XY}$ = ${cov(X, Y)}/{\sigma_{X} \sigma_{Y}}$, $\rho_{XZ}$ = ${cov(X, Z)}/{\sigma_{X} \sigma_{Z}}$, and $\rho_{YZ}$ = ${cov(Y, Z)}/{\sigma_{Y} \sigma_{Z}}$.
Like the Pearson's correlation coefficient, $PrC_{XY}$ also ranges from $-1$ and $1$. 
A higher value indicates a stronger linear relationship between the two fMRI time series after removing the effect of the other regions.
If the value of $PrC_{XY}$ is $0$, it indicates no linear relationship between the two time series.

4) \textbf{Spearman's Correlation} (SC)~\citep{xiao2016using} quantifies the strength of the monotonic relationship between two fMRI time series.
To obtain SC coefficient between two time series $X$ and $Y$, $X$ and $Y$ are first converted to ranked values, followed by:
\begin{equation}
SC_{XY}= 1 - \frac{6 \sum\nolimits d_t^2}{T(T^2 - 1)},
\end{equation}
where $d_t$ is the difference between the ranks of $X$ and $Y$, represented as $d_t = \text{Rank}_X(t)$ - $\text{Rank}_Y(t)$, and $T$ is the number of timepoints. 
$SC_{XY}$ also ranges from $-1$ and $1$, where a value of $-1$ or $1$ implies an exact monotonic relationship between $X$ and $Y$, while a value of $0$ indicates no correlation.

5) \textbf{High-Order Functional Connectivity} (HOFC) \citep{zhang2016topographical} is a measure to examine high-level organization of brain functionality.
Unlike traditional low-order networks (\eg, constructed by Pearson's correlation) that often use functional correlation measures between any pair of brain regions, a HOFC network is constructed based on ``correlation's correlation'', helping characterize high-level inter-region interactions.
Specifically, given the low-order brain network $P\in\mathbb{R}^{N\times N}$, which is constructed based on Pearson's correlation, HOFC between the $i$-th and $j$-th regions can be formulated as:
\begin{equation}
HOFC_{ij} = \frac{\sum\nolimits_{k} (P_{ik} - \overline{P_i})(P_{jk} - \overline{P_j})}{\sqrt{\sum\nolimits_{k} (P_{ik} - \overline{P_i})^2} \sqrt{\sum\nolimits_{k} (P_{jk} - \overline{P_j})^2}},
\end{equation}
where a higher value of $HOFC_{ij}$ represents a stronger high-order relationship between the two regions.

6) \textbf{Sparse Representation} (SR)~\citep{xu2012lr} estimates sparse brain functional networks by introducing an $L1$-regularizer, which can effectively filter out weak or redundant connections in the network.
Mathematically, the objective function for SR-based brain network estimation can be represented as:
\begin{equation}
\min_{\mathbf{W}} \|\mathbf{X} - \mathbf{X}\mathbf{W}\|_{2}^{2} + \lambda \|\mathbf{W}\|_{1},
\label{sr}
\end{equation}
where $W\in\mathbb{R}^{N\times N}$ is expect to represent the original fMRI data $X$ in a sparse manner.
$\lambda$ is the regularization parameter controlling the sparsity level, and a larger value helps produce a sparser brain network.


7) \textbf{Low-rank Representation} (LR)~\citep{recht2010guaranteed} helps construct brain networks in a low-rank learning manner.
Specifically, its optimization 
is formulated as:
\begin{equation}
\min_{\mathbf{W}} \|\mathbf{X} - \mathbf{X}\mathbf{W}\|_{2}^{2} + \lambda \|\mathbf{W}\|_{*},
\end{equation}
where $W\in\mathbb{R}^{N\times N}$ is expected to well represent the original fMRI data $X$ and has a low-rank structure.
$\|W\|_{*}$ denotes the trace norm (\ie, nuclear norm) of $W$, which is the sum of the singular values of $W$.
$\lambda$ is a regularization parameter, with a higher value encouraging a lower-rank brain network.
The low-rank network can represent the original network with fewer components, helping identify the most significant connections and filter out less relevant connections.

It is noted that, besides these methods, users are allowed to use their self-defined algorithms for brain network construction based on fMRI data.
The specific details are elaborated in the \emph{Supplementary Materials}.

\subsubsection{Brain Network Sparsification}
Our toolbox offers two sparsification strategies, allowing users to sparsify the constructed brain networks:

1) \textbf{Sparsity.} This method retains the top $K\%$ values of constructed brain networks while setting the rest to $0$.
In our toolbox, the $K$ is set to $30$ by default.
Compared with the fully connected brain network, the network constructed using ``Sparsity'' can preserve strong connections while removing the weak ones.

2) \textbf{Binarization.} This method follows the same strategy as used in ``Sparsity'', but converts all weighted connections to binary ones.

\subsubsection{Brain Network Visualization}

\begin{figure}[ht]
\setlength{\abovecaptionskip}{-4pt}
\setlength{\belowcaptionskip}{-0pt} \setlength\abovedisplayskip{-0pt}
\setlength\belowdisplayskip{-0pt}
\center
\includegraphics[width= 1.0\linewidth]{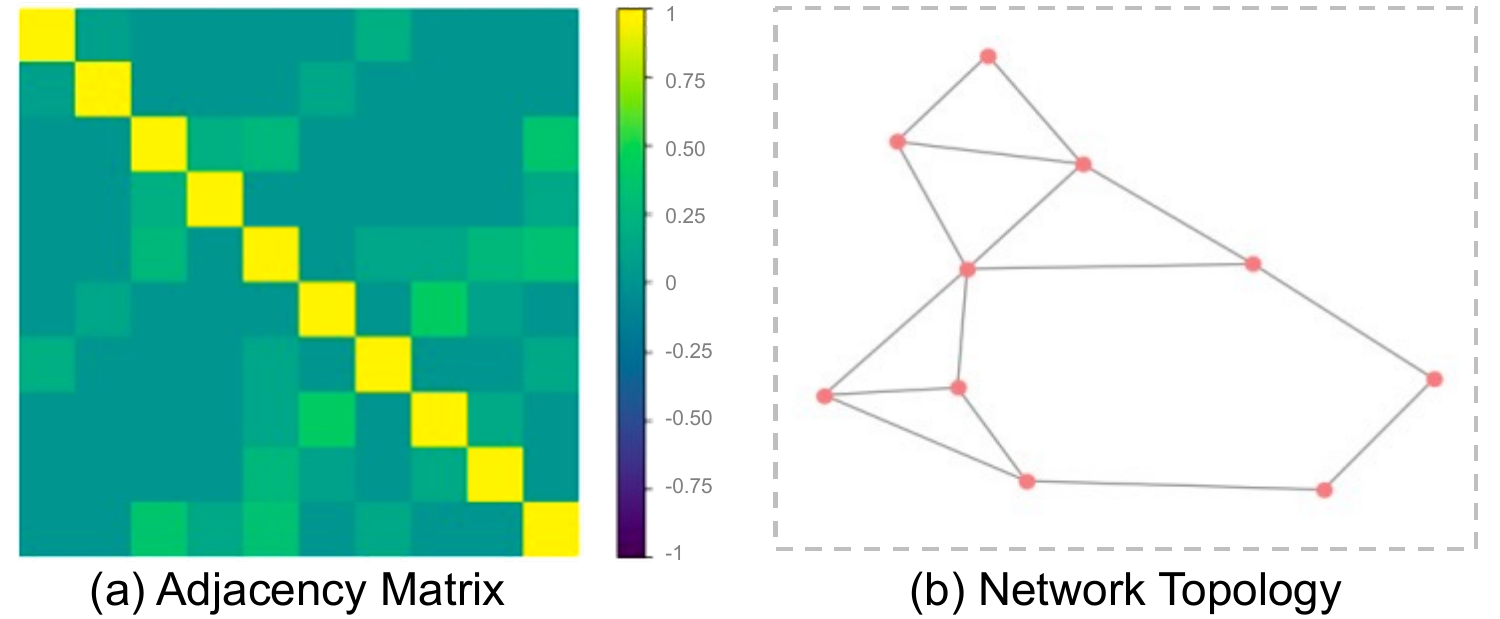}
\caption{Visualization of the constructed brain network.}
\label{fig_network_visualization}
\end{figure}

As a useful addition, this toolbox allows the users to visualize the constructed brain networks from two perspectives, \ie, \emph{adjacency matrix} and \emph{network topology}.

1) \textbf{Adjacency Matrix.}
The constructed brain network can be represented as the adjacency matrix, where each row and column corresponds to a different brain region, and each entry denotes the functional connections between regions.
Here, we use a simulated time series $X\in\mathbb{R}^{N\times T}$ to construct a brain network based on Pearson's correlation, where $N$ = $10$ and $T$ = $30$ denote the numbers of regions and timepoints, respectively.
We retain the top $50\%$ functional connections by setting the Sparsity ratio $K$ to 50. 
The illustration of the adjacency matrix is shown in Fig.~\ref{fig_network_visualization}~(a).

2) \textbf{Network Topology.}
As seen in Fig.~\ref{fig_network_visualization}~(b), we also show the network topology of the constructed brain network based on the same simulated time series.
Here, each graph node represents each brain region, and each graph edge denotes the functional connection between regions.

\subsection{Brain Network Feature Extraction}\label{sec_network_feat_extract}

Based on the constructed brain networks/graphs, we can extract various network features for further fMRI analysis, which are essential for understanding brain functional connectivity patterns. 
Typically, there are two main types of brain network features, \ie, node-based features and graph-based features.
The node-based features are computed for each graph node (\ie, brain region), helping investigate the importance of individual brain regions within the brain network. 
The graph-based features capture global network properties and characterize interactions across brain regions, which provide a holistic view of the entire brain network.
It is noted that our toolbox supports selecting multiple network features simultaneously. 

\subsubsection{Node-based Brain Network Features}
Denote functional relationship between $v_{i}$ and $v_{j}$ as $a_{ij}\in\mathbb{R}^{N\times N}$, where $a_{ij}$ can be binary or weighted.
When $a_{ij}$ is binary, $a_{ij}$=1 if $v_{i}$ and $v_{j}$ connect with each other, otherwise $a_{ij}$=0.
When $a_{ij}$ is weighted, $a_{ij}$ represents the functional connectivities between $v_{i}$ and $v_{j}$. In our work, we introduce six node-based network features, including \emph{node degree}, \emph{node strength}, \emph{local efficiency}, \emph{betweenness centrality}, \emph{eigenvector centrality}, and \emph{clustering coefficient}.

1) \textbf{Node Degree} (ND)~\citep{zegura1996model} calculates the number of edges connected to a node or brain ROI. 
Mathematically, the node degree of $v_{i}$ can be formulated as:
\begin{equation}
ND_i=\sum\nolimits_{j\in N}\textbf{1}(a_{ij}\neq0).
\end{equation}
The brain region with a higher degree represents that it is functionally connected to a greater number of other regions, indicating that it may play a significant role in information communication within the brain network.

2) \textbf{Node Strength} (NS)~\citep{barrat2004architecture} measures a weighted variant of the degree, which is defined as the sum of all neighboring edge weights:
\begin{equation}
NS_i=\sum\nolimits_{j\in N}a_{ij}.
\end{equation}
Node strength is particularly useful in representing brain networks when the connections between brain regions are weighted.
$NS_i$ represents the accumulated functional connectivity between node $v_i$ and the remaining ones.
If $a_{ij}$ is binary, $NS_i$ is similar to $ND_i$.

3) \textbf{Local Efficiency} (LE)~\citep{latora2001efficient} is a measure that quantifies how well a specific brain region communicates with its immediate neighbors, defined as:
\begin{equation}
LE=\frac{1}{n} \sum\nolimits_{i \in V} \frac{\sum\nolimits_{j, h \in V, j \neq i} a_{i j} a_{i h}\left[d_{j h}\left(N_i\right)\right]^{-1}}{k_i\left(k_i-1\right)},
\end{equation}
where $d_{jh}\left(N_i\right)$ is the shortest path length between $v_j$ and $v_h$ that involves only neighboring regions of $v_i$, and $k_i$ is degree of node $v_i$.
A brain region with higher local efficiency indicates that it communicates with its neighbors efficiently.
And a brain network with more such kinds of regions tends to form more densely interconnected clusters, which may facilitate rapid information exchange within the network.


4) \textbf{Betweenness Centrality} (BC)~\citep{freeman2002centrality} is an important measure in functional brain network analysis.
Given a node $v_i$, its betweenness centrality is calculated as the fraction of shortest paths between all pairs of regions in the network that pass through $v_i$, defined as:
\begin{equation}
BC_i=\frac{1}{(N-1)(N-2)} \sum\nolimits\nolimits_{\substack{h, j \in V \\ h \neq j, j \neq i, j \neq i}} \frac{\rho_{h j}(i)}{\rho_{h j}},
\end{equation}
where $\rho_{h j}$ denotes the number of shortest paths between $v_h$ and $v_j$, and $\rho_{h j}(i)$ represents the number of shortest paths between $v_h$ and $v_j$ that pass through $v_i$.
A brain region with a higher BC mean indicates it serves as a critical hub connecting different parts of the network, suggesting that it plays an important role in information transfer within the brain network.

5) \textbf{Eigenvector Centrality} (EC)~\citep{bonacich2007some} quantifies the influence of a brain region based on its connections to other important regions.
Mathematically, the EC of a node $v_i$ is calculated as the eigenvector corresponding to the largest eigenvalue of the adjacency matrix (\ie, $A$) of the network, represented as:
\begin{equation}
EC_i=\frac{1}{\lambda} \sum\nolimits_{j\in V} A_{i j} EC_j,
\end{equation}
where $\lambda$ is the largest eigenvalue of $A$, and $EC_j$ represents the eigenvector centrality of node $v_j$.
A brain region will have a high EC if it is strongly connected with other regions that play significant roles within the network, and thus, EC helps identify influential hub regions in functional networks.

6) \textbf{Clustering Coefficient} (CC)~\citep{watts1998collective} represents the abundance of connected triangles in a brain network.
Given a brain region $v_i$, its CC is computed as the fraction of triangles that exist among the neighbors of $v_i$ out of the all possible number of triangles that could exist among them, defined as follows:
\begin{equation}
CC_i = \frac{1}{n} \sum\nolimits_{i \in V} \frac{2 t_i}{k_i\left(k_i-1\right)},
\end{equation}
where $t_i$ = $\frac{1}{2}\sum\nolimits_{j,h\in V}a_{ij}a_{ih} a_{jh}$ denotes the number of triangles around a node $v_i$, and $k_i$ is degree of node $v_i$.
A node with higher CC indicates that its neighboring regions tend to form tighter interconnected clusters.

\subsubsection{Graph-based Brain Network Features}
We also introduce six graph-based network features, including \emph{density}, \emph{modularity}, \emph{characteristic path length}, \emph{global efficiency}, \emph{assortativity coefficient}, and \emph{transitivity}.

1) \textbf{Density}~\citep{kaiser2011tutorial} quantifies the level of connectivity in the network to measure the percentage of existing connections among all possible connections, defined as:
\begin{equation}
D=\frac{2l}{N(N-1)},
\end{equation}
where $l$ denotes the number of edges in a brain network.
A brain network with higher density suggests that the brain regions are connected more densely.

2) \textbf{Modularity}~\citep{newman2006modularity} quantifies the extent to which a  network can be partitioned into non-overlapping and functionally distinct subnetworks, defined as:
\begin{equation}
M=\frac{1}{2l} \sum\nolimits_{i, j \in V}\left(a_{i j}-\frac{k_i k_j}{2l}\right) \delta_{m_i, m_j}, 
\end{equation}
where $m_i$ is the module containing node $i$, and $\delta_{m_i, m_j}$ equals $1$ if $v_i$ and $v_j$ belong to the same module, otherwise it is $0$.
$k_i$ denotes degree of node $v_i$ and $l$ represents the number of edges.
A brain network with high modularity indicates that it has a modular structure, and brain regions within the same module may share similar functional roles.

3) \textbf{Characteristic Path Length} (CPL)~\citep{watts1998collective} measures the average shortest path length (distance) between all pairs of nodes in the brain network, which can be formulated as:
\begin{equation}
CPL = \frac{1}{N}\sum\nolimits_{i\in V}\frac{\sum\nolimits_{j\in V,j\neq i}d_{ij}}{N-1},
\end{equation}
where $d_{ij}$ is the shortest path between $v_{i}$ and $v_{j}$.
A brain network with higher CPL indicates that information takes longer to transfer across different brain regions, implying lower communication efficiency.
This measure is the most commonly used metric in functional brain network analysis, and many studies have found that its alteration is highly correlated with brain disease disorders, such as Alzheimer's disease~\citep{dai2019disrupted}, Parkinson's disease~\citep{ma2017alteredm}, and epilepsy~\citep{paldino2017metrics}.


4) \textbf{Global Efficiency} (GE)~\citep{latora2001efficient} quantifies the efficiency of information transfer across brain regions in the entire network.
It is defined as the average of the inverse of the shortest path lengths between all pairs of nodes, represented as:
\begin{equation}
GE = \frac{1}{N}\sum\nolimits_{i\in V}\frac{\sum\nolimits_{j\in V,j\neq i}d_{ij}^{-1}}{N-1},
\end{equation}
where $d_{ij}$ denotes the shortest path between $v_{i}$ and $v_{j}$.
A higher global efficiency indicates shorter average path lengths between nodes, suggesting more efficient information transmission across different brain regions.

   


5) \textbf{Assortativity Coefficient} (AC)~\citep{newman2002assortative} quantifies the tendency of nodes with similar degree patterns to connect, which is denoted as:
{\small{
\begin{equation}
AC=\frac{l^{-1} \sum\nolimits_{(i,j) \in E} k_i k_j-\left[l^{-1} \sum\nolimits_{(i,j) \in E} \frac{1}{2}\left(k_i+k_j\right)\right]^2}{l^{-1} \sum\nolimits_{(i,j) \in E} \frac{1}{2}\left(k_i^2+k_j^2\right)-\left[l^{-1} \sum\nolimits_{(i,j) \in E} \frac{1}{2}\left(k_i+k_j\right)\right]^2},
\end{equation}}}
where $k_i$ is degree of node $v_i$, $E$ is the edge set, and $l$ is the number of edges.
AC can provide insights into the organization of connections within the brain network.
A network with larger AC denotes that brain regions tend to connect with other regions of similar degree. 
For instance, regions with high degrees are more likely to associate with other high-degree regions,
which may identify the brain hubs.
Conversely, a brain network with smaller AC suggests that high-degree regions tend to connect with low-degree regions, which implies a more distributed network topology.

6) \textbf{Transitivity}~\citep{newman2003structure} is a measure that quantifies the extent to which connections between neighboring regions are likely to form clusters (\eg, triangles) within the network, which is represented as:
\begin{equation}
T=\frac{\sum\nolimits_{i \in V} 2 t_i}{\sum\nolimits_{i \in V} k_i\left(k_i-1\right)},
\end{equation}
where $t_i$ = $\frac{1}{2}\sum\nolimits_{j,h\in V}a_{ij}a_{ih} a_{jh}$ denotes the number of triangles around a node $v_i$. $k_i$ is degree of node $v_i$.
A higher transitivity indicates that neighboring brain regions are more interconnected, forming local clusters in the network.


\subsection{Artificial Intelligence Model Construction}\label{sec_ai_model_construct}

There is an emerging trend to leverage fMRI data to construct artificial intelligence models for prediction, such as disease diagnosis~\citep{fang2023unsupervised}, age estimation~\citep{lund2022brain}, brain state detection~\citep{wang2018recognizing}.
Many existing toolboxes~\citep{lanka2020malini, zhou2020toolbox, treder2020mvpa, waller2018graphvar} have investigated conventional machine learning models for analyzing fMRI, which mainly rely on hand-crafted features for model construction.
Besides including these conventional models, our toolbox also embeds popular deep learning models for prediction, which are neglected by previous studies. 
The deep learning models can learn data-driven fMRI features guided by downstream tasks, eliminating the need for domain expertise to manually design features.
It is noted that we incorporate a pretraining strategy into each deep learning model, resulting in a backbone encoder with high generalizability that can adapt well to a new dataset/task.
Moreover, we integrate federated learning strategies for each deep learning model, allowing it to be trained using multiple data sites collaboratively while keeping the data decentralized and private.

\subsubsection{Conventional Machine Learning Models}\label{sec_ml_models}

\begin{figure*}[!t]
\setlength{\abovecaptionskip}{-4pt}
\setlength{\belowcaptionskip}{-0pt} \setlength\abovedisplayskip{-0pt}
\setlength\belowdisplayskip{-0pt}
\centering
\includegraphics[width= 0.98\linewidth]{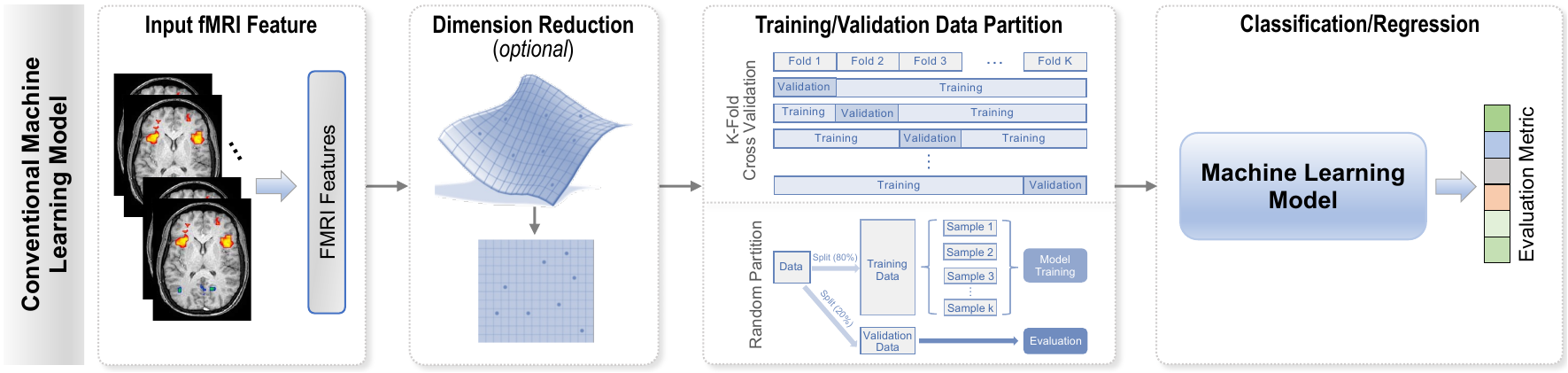}
\caption{Illustration of conventional machine learning framework for fMRI-based prediction.}
\label{fig_machineLearning}
\end{figure*}

Typically, the conventional machine learning models take human-engineered fMRI features as input and then perform model prediction. 
In our toolbox, users can construct machine learning models for both classification and regression tasks, and the whole framework is illustrated in Fig.~\ref{fig_machineLearning}.
Specifically, given the input fMRI features, we can first reduce the feature dimensions, which may eliminate noise and irrelevant features, helping reduce the overfitting issue.
Based on the derived features, we can construct machine learning models for classification/regression.
Several evaluation metrics are provided to verify the effectiveness of the constructed models.
By integrating feature dimension reduction, model construction, and model evaluation into a united framework, our toolbox enables streamlined model prediction based on fMRI features.
The main components in this machine learning framework are detailed as follows.

\textbf{Techniques for Feature Dimension Reduction.}
Since the input fMRI features may be high-dimensional, our toolbox allows users to reduce feature dimensions before performing classification/regression.
Three popular dimension reduction techniques are embedded in ACTION, including:
\begin{itemize}
    \item {Principal Component Analysis} (PCA)~\citep{wold1987principal}, which reduces dimension by keeping the features (\ie, principal components) that contribute most to data variance;
    \item {Independent Component Analysis} (ICA)~\citep{hyvarinen1997fast}, which performs dimension reduction by separating the features into a set of additive and independent non-Gaussian components;
    \item {Canonical Correlation Analysis} (CCA)~\citep{hotelling1992relations}, which aims to extract the most informative dimensions by identifying linear combinations that maximize the correlation of the input features.
\end{itemize}

\textbf{Model Description.}
This toolbox integrates several popular machine learning models to analyze fMRI, including:
\begin{itemize}
    \item Support Vector Machine (SVM)~\citep{hearst1998support}, which performs prediction by finding the hyperplane that best separates different classes of data with maximum margin;
    \item Random Forest (RF)~\citep{breiman2001random}, which is an ensemble learning method that builds multiple decision trees and merges them to obtain prediction; 
    \item Extreme Gradient Boosting (XGBoost)~\citep{chen2016xgboost}, which is an optimized ensemble learning method that combines decision trees and gradient boosting for prediction;
    \item K-Nearest Neighbors (KNN)~\citep{fix1985discriminatory}, which performs prediction based on the majority class or average of its nearest neighbors in the feature space.
\end{itemize}

These methods are suited for classification and regression tasks, enabling users to address different downstream tasks.
Users can design and use their self-defined models for classification/regression, with details given in \emph{Supplementary Materials}. 
Users can utilize fMRI features derived from ACTION (\eg, ``brain network feature extraction'' module), and also take features generated by themselves.

\begin{figure*}[!t]
\center
\includegraphics[width= 0.98\linewidth]{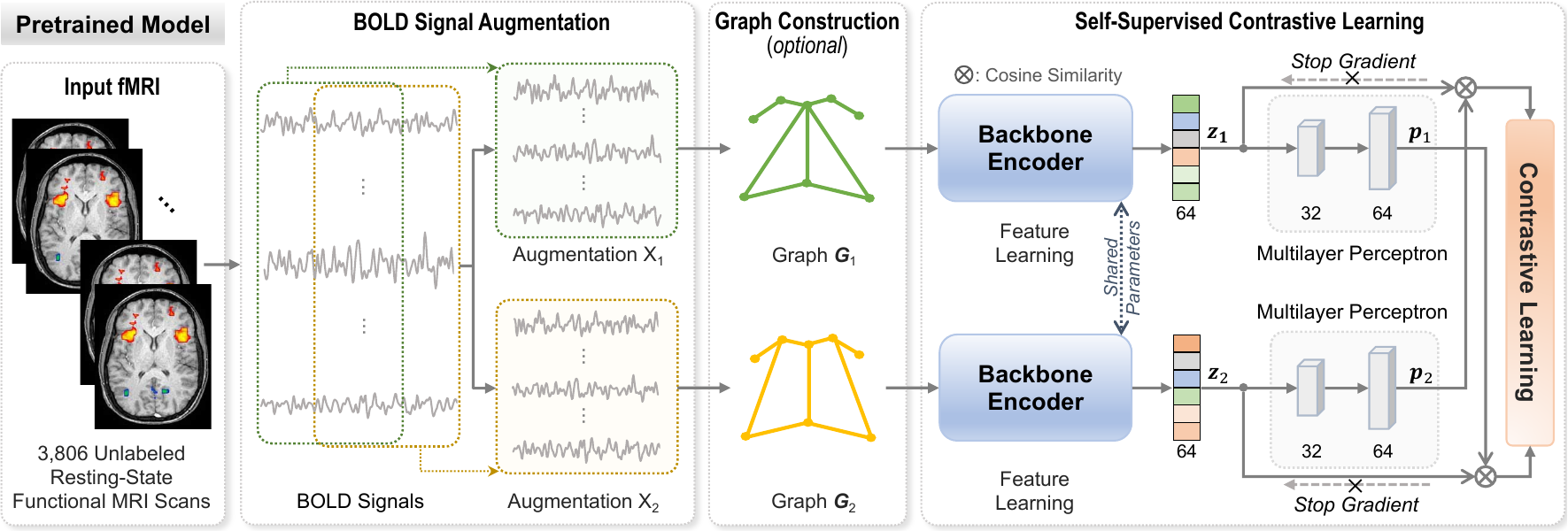}
\caption{Illustration of self-supervised contrastive learning framework for pretraining deep learning models with fMRI.}
\label{fig_DeepLearning}
\end{figure*}

\textbf{Data Partition Strategies.}
We divide the input data into training and validation sets to evaluate the performance of the machine learning models. 
The training set is used to train the model while the validation set is used to assess model performance based on the trained model. 
Our toolbox provides two different data partition strategies:
 
\begin{itemize}
\item $K$-fold Cross Validation: This strategy divides the input data into $K$ distinct subsets.
During model training, one subset is treated as the validation set to evaluate model performance, while the remaining $K$-1 subsets are used for model training. 
This process is repeated $K$ times, with each $K$ subset serving as the validation set once. 
The final prediction is derived by averaging the results from these $K$-fold validation results.
This data partition strategy ensures the model is validated across all samples. 
\item Random Partition: This strategy randomly divides the input data according to a specified partition ratio $R$\%.
That is, $R$\% data samples are used for model training while the remaining data are used for validation.
\end{itemize}

\textbf{Evaluation Metrics.}
Our toolbox provides several metrics for model evaluation.
For classification tasks, seven metrics are used: area under the receiver operating characteristic curve (AUC), accuracy, balanced accuracy, F1-score, sensitivity, specificity, and precision.
It also allows users to plot confusion matrices and AUC to visualize classification results.
For regression tasks, we evaluate the model using mean absolute error, mean squared error, and concordance correlation coefficient in this toolbox.  

\subsubsection{Deep Learning Models}\label{sec_deep_models}
In ACTION, we incorporate ten popular deep learning methods for computer-aided fMRI analysis.
Moreover, for each method, our toolbox constructs a pretrained deep learning model based on large-scale unlabeled fMRI data in a self-supervised manner.
As the pretrained models are built based on large and diverse fMRI data, they are expected to capture more general fMRI features, helping improve model performance in downstream tasks.
In addition, the pretrained models can be used to finetune downstream tasks with limited sample sizes, facilitating effective knowledge transfer and reducing the model overfitting issue.

The framework for constructing a pretrained deep learning model is illustrated in Fig.~\ref{fig_DeepLearning}.
For training the model, we use 3,806 resting-state fMRI scans from three public datasets, including Autism Brain Imaging Data Exchange (ABIDE) initiative\footnote{\url{http://fcon_1000.projects.nitrc.org/indi/abide}}, 
REST-meta-MDD Consortium\footnote{\url{http://rfmri.org/REST-meta-MDD}}, 
and ADHD-200\footnote{\url{http://fcon_1000.projects.nitrc.org/indi/adhd200/}}. 
The selection criteria of these fMRI scans and their ID information are given in \emph{Supplementary Materials}. 
As shown in Fig.~\ref{fig_DeepLearning}, the framework contains three main components, including fMRI BOLD signal augmentation, graph construction, and self-supervised contrastive learning.
With 3,806 fMRI time series as input, we first perform data augmentation using a slicing strategy, yielding two augmented signals $X_{1}$ and $X_{2}$.
Here, $X_{1}$ and $X_{2}$ are obtained by segmenting the first 90\% and the last 90\% of full-length signals, respectively.
Then, they are fed into a graph construction module for fMRI feature learning.
This step is optional, depending on the specific architecture of deep learning methods.
After that, $X_{i}$ or $G(X_{i})$ ($i$=1 or 2) are input to two shared backbone encoders for feature learning, resulting in fMRI representations $z_1$ and $z_2$, respectively.
Based on $z_1$ and $z_2$, we leverage two multilayer perceptron (MLP) layers to abstract higher-level feature representations $p_{1}$ and $p_{2}$. 
Inspired by SimSiam~\citep{chen2021exploring}, the pretrained model is optimized by maximizing the agreement between two augmented features based on contrastive learning.
The main idea is to enforce the consistency of augmented features from the same fMRI scan.
Mathematically, this optimization process is as follows:
\begin{equation}
\mathbb{L}_{C}=\Phi(\psi(z_{1}),p_2)+\Phi(\psi(z_{2}),p_1),
\end{equation}
where $\Phi$ denotes the negative cosine similarity and $\psi$ represents the stop-gradient operation, which enables the stability of the training process~\citep{chen2021exploring}.
After obtaining the pretrained backbone encoder, we can finetune it based on users' downstream tasks, helping adapt the model to a new fMRI dataset.
The specific finetuning process for each deep learning method is detailed in our 
\href{https://github.com/mxliu/ACTION-Software-for-Functional-MRI-Analysis/tree/main/Software}{open-source code}.

In the following, we detail the ten deep learning methods included in ACTION for fMRI analysis.

1) {Transformer}~\citep{vaswani2017attention}, a deep learning model based on self-attention, has become foundational for processing sequential data.
For fMRI analysis, we first construct the brain network based on fMRI time series via Pearson's correlation.
We then leverage self-attention to dynamically weigh the importance of different regions in the network, capturing dependencies among brain regions.

2) {Graph Convolutional Network} (GCN)~\citep{kipf2016semi} is a powerful graph neural network designed specifically for handling graph-structured data, \eg, brain networks.
In our case, we also use Pearson's correlation to construct the brain network/graph.
Then, we utilize two stacked graph convolutional layers to update and aggregate the representations of each graph node/brain region, yielding a hierarchical representation of the brain network.

3) {Graph Attention Network} (GAT)~\citep{velivckovic2017graph} extends the concept of GCN~\citep{kipf2016semi} by introducing the mechanism of attention.
Unlike GCN which treats contributions of all regions equally, GAT employs a learnable attention mask that dynamically assigns different weights to each brain region, enabling the model to focus more on task-relevant information.
    
4) {Graph Isomorphism Network} (GIN)~\citep{kim2020understanding} achieves maximum discriminative power among graph neural networks by generalizing the Weisfeiler-Lehman (WL) test.
Like GCN~\citep{kipf2016semi}, with the brain network as input, 
we stack two GIN layers for fMRI feature learning, followed by a pooling operation to generate a graph representation.

5) {Graph Sample and Aggregate} (GraphSAGE)~\citep{hamilton2017inductive} is a method designed for analyzing graph-structured data. 
With the brain network as input, GraphSAGE learns node representations by sampling and aggregating information from its local neighborhood.
To abstract fMRI features, two GraphSAGE layers are leveraged, followed by a pooling operation.

6) {Brain Network Convolutional Neural Network} (BrainNetCNN)~\citep{kawahara2017brainnetcnn} is specially designed for brain network analysis, consisting of 3 convolutional filters (edge-to-edge, edge-to-node, and node-to-graph) to capture topological structure information of brain networks.

7) {Brain Graph Neural Network} (BrainGNN)~\citep{li2021braingnn} is a graph neural network designed for analyzing fMRI and detecting neurological biomarkers.
With the brain network as input, BrainGNN uses two node-level graph convolutional layers to learn the node representations, capturing topological and functional patterns from fMRI data.

8) {Spatio-Temporal Graph Convolutional Network} (STGCN)
~\citep{gadgil2020spatio} is designed to jointly extract spatial and temporal features from fMRI times series via spatiotemporal graph convolution units (GCUs).
Here, we stack two GCUs to model spatiotemporal patterns, and then generate a graph representation via a pooling operation.

9) {Spatio-Temporal Attention Graph Isomorphism Network} (STAGIN)~\citep{kim2021learning} is designed to model fMRI dynamics using spatiotemporal attention. 
Specifically, it first partitions the fMRI time series using a sliding-window scheme and employs GIN~\citep{kim2020understanding} to aggregate node features in each window.
Then, a Transformer~\citep{vaswani2017attention} is leveraged to capture temporal attention across different windows to characterize fMRI dynamics, resulting in a spatiotemporal graph representation.

10) {Modularity-constrained Graph Neural Network} (MGNN)~\citep{wang2024leveraging} is specially designed to learn spatiotemporal dynamic representations of fMRI.
MGNN provides a novel scheme to incorporate brain modularity to learn fMRI features, which encourages node representations within the same module to be similar.
A graph-level feature representation is then generated via a pooling operation.

Additionally, our toolbox offers five federated learning strategies, which allow the model to be trained across many decentralized data sites, facilitating multi-site fMRI studies.
These strategies are introduced in the following. 

1) {Federated Averaging} (\textbf{FedAvg})~\citep{mcmahan2017communication} is a widely-used distributed learning paradigm.
Specifically, the local sites first copy the global model parameters, which are initialized randomly or by our pretrained model.
These sites independently calculate gradients based on their local data and send the gradients' updates to the central server.
The server then updates the global model by averaging these updates, and the updated model is sent back to each site for the next training round.

2) {Federated Proximal} (\textbf{FedProx})~\citep{li2020federated}, which refines FedAvg by addressing cross-site heterogeneity.
Specifically, like FedAvg, FedProx first obtains a global model by averaging parameters received from each local site.
Then, it mitigates the bias between global and local model parameters by introducing a regularization (\ie, proximal term) to the optimization objective for each site, helping eliminate parameter drift.
    
3) {Model-Contrastive Federated Learning} (\textbf{MOON})~\citep{li2021model}, which is designed based on FedAvg, where a global model is first constructed by averaging local sites' parameters.
Then, it maximizes the similarity between representations learned by the local and global models, and minimizes the similarity between representations of the local model in the current training round and that in the previous round, helping correct the local training.

4) {Local Global Federated Avgeraging} (\textbf{LGFedAvg}) \citep{liang2020think}, which captures compact local representations on each site and a global model across all sites.
Specifically, it sends the parameters of the last layer in the local site to the central server for aggregation while other parameters remain at each local site.
In this way, the number of communication parameters can be much smaller than other federated learning algorithms, \eg, FedAvg.

5) {Personalized Federated Learning with Moreau Envelopes} (\textbf{pFedMe})~\citep{t2020personalized}, which aims to address statistical diversity among different sites.
It uses Moreau envelopes as local sites' regularized loss functions, helping decouple the optimization process of the local model from the global model learning process.
Thus, the global model can be utilized to optimize the local model.

\begin{table*}[tp]
\setlength{\abovecaptionskip}{0pt}
\setlength{\belowcaptionskip}{-2pt}
\setlength{\abovedisplayskip}{-2pt}
\setlength{\belowdisplayskip}{-2pt}
\setlength\tabcolsep{2pt}
\renewcommand{\arraystretch}{0.8}
\caption{Demographic characteristics of the studied subjects of three sites (\ie, NYU, UM, and UCLA) from the public ABIDE cohort~\citep{craddock2013neuro}.
ASD: autism spectrum disorder; 
HC: healthy control; 
M/F: Male/Female;
std: standard deviation.}
\label{tab_demographic_info}
\centering
\resizebox{0.9\textwidth}{!}{
\begin{tabular*}{1.05\textwidth}{@{\extracolsep{\fill}} l|ccc|ccc|ccc}
    \toprule
    \multirow{2}{*}{~Group}&
    \multicolumn{2}{c}{NYU} & \phantom{ab} & \multicolumn{2}{c}{UM} & \phantom{ab} & \multicolumn{2}{c}{UCLA}\\
    \cmidrule(lr){2-3} \cmidrule(lr){5-6} \cmidrule(lr){8-9}
    & ASD & HC && ASD & HC && ASD & HC &\\
    \midrule
    ~Subject No. & 50 & 50 && 40 & 40 && 30 & 30 \\
    ~Gender (M/F) & 42/8 & 39/11 && 34/6 & 30/10 && 28/2 & 25/5 \\
    ~Age (mean$\pm$std)& 12.39±5.51 & 15.12±6.58 && 13.55±2.32 & 14.46±2.89 && 12.86±2.45 & 13.09±2.06 \\
    \bottomrule
\end{tabular*}}
\end{table*}

\begin{table*}
\setlength{\belowdisplayskip}{-0pt}
\setlength{\abovedisplayskip}{-0pt}
\setlength{\abovecaptionskip}{-0pt}
\setlength{\belowcaptionskip}{-0pt}
\scriptsize
\setlength\tabcolsep{3pt}
\renewcommand\arraystretch{1}
\caption{Results of machine learning models in ASD vs. HC classification on NYU.
ASD: autism spectrum disorder;
HC: healthy control.}
\label{tab_ml_models}
\centering
\vspace{2mm}
\begin{tabular*}{1\textwidth}{@{\extracolsep{\fill}} l|ccccccc}
    \toprule
    ~Model & AUC (\%) & Accuracy (\%) & Balanced Accuracy (\%) & F1-score (\%) & Sensitivity (\%) & Specificity (\%) & Precision (\%) \\
    \midrule
    ~SVM & 89.88±2.25 & 79.00±8.60 & 79.94±8.20 & 78.88±10.64 & 83.91±6.74 & 75.97±13.97 & 76.21±16.42 \\
    ~RF & 85.68±8.46 & 75.00±12.25 & 75.83±12.26 & 73.77±11.26 & 70.83±19.36 & 80.83±13.33 & 81.33±11.62 \\
    ~XGBoost & 83.92±7.45 & 72.00±10.30 & 72.76±10.28 & 72.23±11.24 & 76.95±16.36 & 68.57±9.93 & 69.81±11.56 \\
    ~KNN & 77.35±10.64 & 63.00±9.80 & 63.57±5.07 & 71.38±9.44 & 96.35±4.64 & 30.80±8.54 & 57.79±12.55 \\
    \bottomrule
\end{tabular*}
\end{table*}

\section{Empirical Evaluation}
\label{sec_experiment_validation}

\subsection{Materials and Data Processing} 
\subsubsection{Datasets}
We employ a real fMRI dataset (called NYU) for model evaluation.
This dataset consists of 50 patients diagnosed with autism spectrum disorder (ASD) and 50 healthy controls (HCs).
All subjects are randomly sampled from the NYU site of ABIDE\footnote{\url{http://fcon_1000.projects.nitrc.org/indi/abide}}. 
We utilize this dataset to explore the diagnostic capabilities of all introduced models, \ie, classifying ASD patients from HCs.
Moreover, to evaluate the effectiveness of federated learning strategies, we include two more sites from ABIDE, \ie, UM and UCLA, to help train a global model and evaluate diagnostic performance in each site.
The demographic characteristics of all studied subjects are shown in Table~\ref{tab_demographic_info}, while subject IDs are reported in \emph{Supplementary Materials}.

\subsubsection{Data Preprocessing}
A standardized fMRI preprocessing pipeline based on Data Processing Assistant for Resting-State fMRI (DPARSF)~\citep{yan2010dparsf} is utilized for preprocessing all fMRI scans, including discarding the first 10 volumes, slice timing correction, head motion estimation, bandpass filtering, regression of nuisance covariates, co-registration between T1-weighted images and mean functional images, and transformations from individual native space to the Montreal Neurological Institute (MNI) template space. 
In this work, we use the Automated Anatomical Labeling (AAL) atlas with $N = 116$ ROIs for brain ROI parcellation, resulting in regional mean fMRI time series for each subject.

\subsection{Evaluation of Machine Learning Models}
\begin{figure*}[!t]
\setlength{\abovecaptionskip}{-8pt}
\setlength{\belowcaptionskip}{-0pt} 
\setlength\abovedisplayskip{-0pt}
\setlength\belowdisplayskip{-0pt}
\center
\includegraphics[width= 1.0\linewidth]{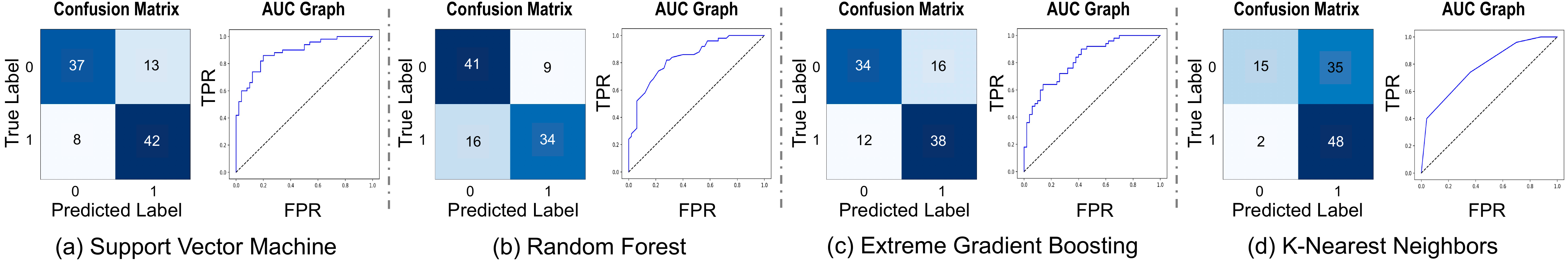}
\caption{Demonstration of the confusion matrix and AUC graph generated by four conventional machine learning methods for ASD diagnosis on NYU.
AUC: area under the receiver operating characteristic curve; 
TPR: true positive rate; 
FPR: false positive rate; 
1: patients with autism spectrum disorder (ASD);
0: healthy controls.}
\label{fig_confusematrix_auc}
\end{figure*}

We first validate the performance of conventional machine learning models introduced in Section~\ref{sec_ml_models} based on fMRI data from NYU.
Specifically, given the fMRI time series data, we first use Pearson's correlation to construct a functional brain network for each subject.
Then, we flatten the upper triangle elements of the network and convert them into a vectorized representation.
For all experiments, we utilize PCA to reduce the feature dimension to 20, and we choose 5-fold cross-validation for model training.
After inputting fMRI features and corresponding diagnostic labels into our toolbox, we can easily obtain classification results for each model, shown in Table~\ref{tab_ml_models}.
It can be seen from Table~\ref{tab_ml_models} that these machine learning models yield promising results in classifying ASD patients from HCs, indicating their effectiveness and efficacy.
Additionally, our toolbox provides functionality for plotting the confusion matrix and AUC graph, enabling users to visualize prediction results. 
Fig.~\ref{fig_confusematrix_auc} presents these results for the conventional machine learning models included in ACTION.

\begin{table*}[!t]
\setlength{\belowdisplayskip}{-8pt}
\setlength{\abovedisplayskip}{-8pt}
\setlength{\abovecaptionskip}{-0pt}
\setlength{\belowcaptionskip}{-0pt}
\scriptsize
\setlength\tabcolsep{3pt}
\renewcommand\arraystretch{0.9}
\caption{Results of deep learning models in ASD vs. HC classification on NYU.
ASD: autism spectrum disorder;
HC: healthy control.}
\label{tab_deep_models}
\centering
\vspace{2mm}
\begin{tabular*}{1\textwidth}{@{\extracolsep{\fill}} l|ccccccc}
    \toprule
    Model & AUC (\%) & Accuracy (\%) & Balanced Accuracy (\%) & F1-score (\%) & Sensitivity (\%) & Specificity (\%) & Precision (\%) \\
    \midrule
    Transformer & 76.00±14.36 & 60.00±16.96 & 63.13±15.91 & 67.49±12.85 & 83.16±15.32 & 43.09±36.34 & 60.68±23.02 \\
    GCN & 74.61±11.33 & 59.00±14.32 & 62.69±11.74 & 58.99±6.57 & 61.03±25.02 & 64.36±43.53 & 73.00±27.75 \\
    GAT & 77.87±10.96 & 67.00±9.75 & 68.53±10.89 & 65.54±12.98 & 66.12±17.60 & 70.94±24.31 & 69.56±20.30 \\
    GIN & 72.17±12.56 & 61.00±12.94 & 64.45±12.09 & 64.21±7.04 & 70.32±18.85 & 58.58±37.38 & 67.23±24.61 \\
    GraphSAGE & 72.12±17.57 & 66.00±13.87 & 67.37±13.02 & 62.10±26.94 & 72.02±35.84 & 62.72±29.00 & 71.56±22.84 \\
    BrainNetCNN & 82.92±14.00 & 79.00±9.62 & 77.94±9.29 & 77.04±13.50 & 77.07±22.71 & 78.80±13.43 & 79.88±7.86 \\
    BrainGNN & 87.09±3.98 & 75.00±8.66 & 75.56±11.16 & 69.15±20.50 & 65.85±27.43 & 85.27±12.10 & 79.57±13.37 \\
    STGCN & 61.25±9.95 & 60.00±6.12 & 58.75±4.82 & 61.61±6.87 & 67.18±20.44 & 50.33±28.21 & 60.21±8.51 \\
    STAGIN & 87.51±11.82 & 79.00±8.22 & 80.99±7.60 & 80.16±7.52 & 86.77±20.34 & 75.21±18.37 & 78.88±12.74 \\
    MGNN & 78.16±5.70 & 73.00±7.58 & 72.95±7.57 & 73.18±10.71 & 77.72±8.92 & 68.18±7.78 & 69.92±14.48 \\
    \bottomrule
\end{tabular*}
\end{table*}

\subsection{Evaluation of Deep Learning Models}

In the first group of experiments, we validate the performance of ten deep learning models for ASD diagnosis on fMRI from NYU using a 5-fold cross-validation strategy.
All models are initialized using the pretrained deep learning models and finetuned on the training set of NYU, with classification results shown in Table~\ref{tab_deep_models}. 
From Table~\ref{tab_deep_models}, we can observe that these deep learning models achieve satisfactory performance in ASD identification.

In the second group of experiments, we evaluate the effectiveness of federated learning strategies introduced in our toolbox. 
Here, we employ the pretrained GCN as the baseline model, based on which we apply different strategies.
The classification results for ASD diagnosis of five federated learning strategies are reported in Table~\ref{tab_federated_learning}.
Moreover, we include two non-federated learning methods (\ie, Single and Mix) for comparison.
Specifically, the ``Single'' method only uses data from a single site for model training and test via 5-fold cross-validation, without performing knowledge transfer/sharing among the individual sites. 
The ``Mix'' method uses all data pooled from all sites. 
The results of the ``Single'' method in ``Average Results'' are derived as follows. We first obtain prediction results for each site and then concatenate these results from all three sites. Then, with corresponding ground-truth labels, we can perform model prediction and obtain the results. 
A 5-fold cross-validation strategy is employed in these seven competing methods. 

It can be seen from Table~\ref{tab_federated_learning} that the federated learning models generally outperform the ``Single'' method. 
The underlying reason may be that federated learning enables capturing diverse fMRI features from multiple data sites, which can help enhance model generalization and thus yield better prediction results.
In addition, these models also show superior classification results compared with the ``Mix'' method.
The possible reason may be that federated learning enables multiple sites to train models collaboratively, which allows each site to leverage complementary knowledge from other sites, thus enhancing classification performance.


\subsection{Limitation and Future Work}
Several limitations need to be addressed in the future. 
\emph{First}, current toolbox investigates fMRI data augmentation strategies from time series and graph perspectives.
Benefiting from promising prospects of generative models in data augmentation, future work will explore innovative generative methods (\eg, diffusion models~\citep{yang2023diffusion}) to enhance data diversity and scale.
\emph{Second}, the existing toolbox limits users to training deep learning models using only their local computing resources. 
In the future, we plan to develop a cloud computation platform, which empowers users with limited computation resources to engage in deep learning fMRI analysis.
\emph{In addition}, although our toolbox offers a user-friendly interface with graphical controls and visualizations, it is constrained by compatibility with specific package versions.
We intend to address this issue by creating a Docker container that encapsulates all necessary environments and dependencies in the future.
\emph{Lastly}, current work only supports the construction of AI-based models individually without leveraging multiple models that may capture complementary fMRI patterns.
It is interesting to incorporate advanced ensemble learning algorithms~\citep{yang2023survey} into this toolbox to further boost its utility. 


\section{Conclusion}\label{sec_conclusion}
This paper introduces a Python-based cross-platform toolbox, called ACTION, for computer-aided functional MRI analysis. 
The ACTION consists of four components: \ie, fMRI data augmentation, brain network construction, brain network feature extraction, and artificial intelligent model construction.
It incorporates state-of-the-art fMRI data augmentation strategies and deep learning models. 
Moreover, federated learning strategies are embedded in our toolbox to help users implement their multisite fMRI studies without centralized data storage and computation. 
Experiments on three fMRI sites suggest the effectiveness and user-friendliness of ACTION.
We hope our ACTION can benefit researchers in analyzing fMRI more efficiently.

\begin{table*}[!t]
\setlength{\belowdisplayskip}{-0pt}
\setlength{\abovedisplayskip}{-0pt}
\setlength{\abovecaptionskip}{-0pt}
\setlength{\belowcaptionskip}{-0pt}
\scriptsize
\setlength\tabcolsep{3pt}
\renewcommand\arraystretch{0.8}
\caption{Results of federated learning models and baselines in ASD vs. HC classification on three sites (\ie, NYU, UM, and UCLA). 
The average results across three sites are also provided.
ASD: autism spectrum disorder;
HC: healthy control.}
\label{tab_federated_learning}
\centering
\vspace{2mm}
\begin{tabular*}{1\textwidth}{@{\extracolsep{\fill}} l|ccccccc}
    \toprule
    Results on  NYU Site & AUC (\%) & Accuracy (\%) & Balanced Accuracy (\%) & F1-score (\%) & Sensitivity (\%) & Specificity (\%) & Precision (\%) \\
    \midrule
    Single & 65.6±11.0 & 65.0±11.4 & 65.0±11.4 & 70.5±8.1 & 84.0±16.0 & 46.0±28.7 & 60.9±11.3 \\
    FedAvg & 75.2±9.1 & 69.0±10.2 & 69.0±10.2 & 68.0±9.9 & 66.0±9.4 & 72.0±13.2 & 70.2±12.1 \\
    FedProx & 75.6±12.9 & 69.0±9.1 & 69.0±9.1 & 68.7±8.3 & 68.0±11.7 & 70.0±16.7 & 69.4±12.1 \\
    MOON & 74.3±9.2 & 72.0±8.3 & 72.0±8.3 & 71.4±15.2 & 70.0±9.3 & 74.0±11.9 & 72.9±11.7 \\
    LGFedAvg & 75.8±10.7 & 73.0±10.7 & 73.0±10.7 & 73.8±11.4 & 76.0±13.6 & 70.0±9.4 & 71.7±11.1 \\
    pFedMe & 75.4±8.2 & 71.0±9.5 & 71.0±9.5 & 71.3±9.6 & 72.0±10.1 & 70.0±11.7 & 70.6±13.1 \\
    \midrule
    Results on UM Site & AUC (\%) & Accuracy (\%) & Balanced Accuracy (\%) & F1-score (\%) & Sensitivity (\%) & Specificity (\%) & Precision (\%) \\
    \midrule
    Single & 73.4±19.0 & 61.7±6.7 & 61.7±6.7 & 41.0±17.7 & 26.7±26.7 & 96.7±6.7 & 88.9±8.0 \\
    FedAvg & 54.1±12.6 & 56.7±6.2 & 56.7±6.2 & 31.6±19.1 & 20.0±25.5 & 93.3±8.2 & 75.0±27.4 \\
    FedProx & 63.3±18.2 & 60.0±14.3 & 60.0±14.3 & 60.0±22.3 & 60.0±25.8 & 60.0±17.0 & 60.0±16.3 \\
    MOON & 62.6±13.8 & 61.7±11.3 & 61.7±11.3 & 62.3±15.9 & 63.3±19.4 & 60.0±22.6 & 61.3±11.9 \\
    LGFedAvg & 74.8±12.0 & 63.3±12.5 & 63.3±12.5 & 62.1±13.5 & 60.0±16.3 & 66.7±14.9 & 64.3±15.1 \\
    pFedMe & 63.3±15.1 & 60.0±13.3 & 60.0±13.3 & 55.6±19.8 & 50.0±24.5 & 70.0±12.5 & 62.5±18.0 \\
    \midrule
    Results on UCLA Site & AUC (\%) & Accuracy (\%) & Balanced Accuracy (\%) & F1-score (\%) & Sensitivity (\%) & Specificity (\%) & Precision (\%) \\
    \midrule
    Single & 64.3±14.2 & 63.7±12.1 & 63.7±12.1 & 72.9±7.3 & 97.5±30.0 & 30.0±23.2 & 58.2±8.8 \\
    FedAvg & 69.1±17.8 & 63.8±12.1 & 63.8±12.1 & 54.0±15.8 & 42.5±14.2 & 85.0±12.2 & 73.9±9.8 \\
    FedProx & 67.2±17.8 & 65.0±18.8 & 65.0±18.8 & 69.6±15.9 & 80.0±21.5 & 50.0±16.2 & 61.5±8.8 \\
    MOON & 71.5±15.7 & 63.8±12.7 & 63.8±12.7 & 69.5±10.3 & 82.5±21.5 & 45.0±20.3 & 60.0±10.2 \\
    LGFedAvg & 66.0±15.9 & 62.5±11.1 & 62.5±11.1 & 58.3±7.7 & 52.5±6.1 & 72.5±21.5 & 65.6±19.1 \\
    pFedMe & 67.6±17.0 & 62.5±16.3 & 62.5±16.3 & 65.9±15.5 & 72.5±17.6 & 52.5±16.5 & 60.4±12.8 \\
    \midrule
    Average Results & AUC (\%) & Accuracy (\%) & Balanced Accuracy (\%) & F1-score (\%) & Sensitivity (\%) & Specificity (\%) & Precision (\%) \\
    \midrule
    Single & 64.3±6.4 & 63.7±4.3 & 63.7±4.3 & 67.2±4.5 & 74.2±8.5 & 53.3±4.9 & 61.3±3.5 \\
    Mix & 63.6±3.1 & 60.4±2.6 & 60.4±2.6 & 56.2±11.6 & 50.8±21.5 & 70.0±20.5 & 62.9±10.7 \\
    FedAvg & 66.4±6.0 & 64.2±4.8 & 64.2±4.8 & 56.6±6.1 & 46.7±19.1 & 81.7±7.3 & 71.8±8.7 \\
    FedProx & 69.7±7.3 & 65.4±9.4 & 65.4±9.4 & 66.9±8.1 & 70.0±7.5 & 60.8±15.0 & 64.1±9.8 \\
    MOON & 70.8±9.4 & 66.7±9.6 & 66.7±9.6 & 68.5±9.0 & 72.5±11.0 & 60.8±11.1 & 64.9±8.4 \\
    LGFedAvg & 71.9±8.6 & 67.1±8.3 & 67.1±8.3 & 66.1±7.4 & 64.2±8.5 & 70.0±12.2 & 68.1±8.9 \\
    pFedMe & 69.8±7.6 & 65.4±8.2 & 65.4±8.2 & 65.8±8.1 & 66.7±11.4 & 64.2±11.4 & 65.0±8.4 \\
    \bottomrule
\end{tabular*}
\end{table*}

\section*{Declarations of Interest}
The authors declare that they have no known competing financial interests or personal relationships that could have appeared to influence the work reported in this paper. 

\section*{Acknowledgement}\label{sec_acknowledge}
Part of the data used in this work are from the ABIDE initiative, REST-meta-MDD project, and ADHD-200 Sample initiative. 
The investigators of ABIDE, REST-meta-MDD, and ADHD-200 provide the data but are not involved in data processing, analysis, toolbox development, and writing.

\bibliographystyle{cas-model2-names}

\bibliography{Toolbox_ACTION}

\end{sloppypar}
\end{document}